# Knowledge Consilience: One Culture, Two Cultures or Many Cultures?

Nick Zhang

**The hostility between the two cultures, scientific and literary, was framed by C.P. Snow in 1959[1] and later by others[2,3,4]. The scientific culture is nowadays often identified with STEM (Science, Technology, Engineering and Mathematics) whereas the literary culture generally refers to humanities and social sciences. Wilson expressed the wish for the unity of knowledge[5]. We put forward the notions of knowledge distance and knowledge consilience threshold to quantitatively measure distance and coupling process between different branches of knowledge. Our findings suggest that the gulf between the two cultures is widening.**

Microsoft Academic Graph (MAG)[6] contains more than 240 million academic publications as of 2020. A hierarchical ontology, FOS ("Field of Study"), is built by using concept detection and taxonomy learning techniques[7]. There are 19 top fields, such as, 'Mathematics', 'Engineering', 'Political Science' etc. A top field may contain secondary subfields. 'Mathematics', for example, contains 'Algebra', which in turn contains 'Abstract Algebra'. In MAG, a paper is a node and a field is a collection of papers belonging to it. The inbound and outbound links of a collection are the sums of citations to and from all papers it contains as shown in Fig. 1.

**Knowledge distance.**

Simple citation count cannot adequately characterize the relationships between academic disciplines. Google similarity distance[8] measures semantic distance between different nodes in a network. It was used to model the semantic relatedness between different articles in Wikipedia[9]. Experiments have shown that it consistently outperforms other metrics in accuracy and robustness, with additional advantage of being agnostic to data size, therefore resilient to citation-inflation[10].

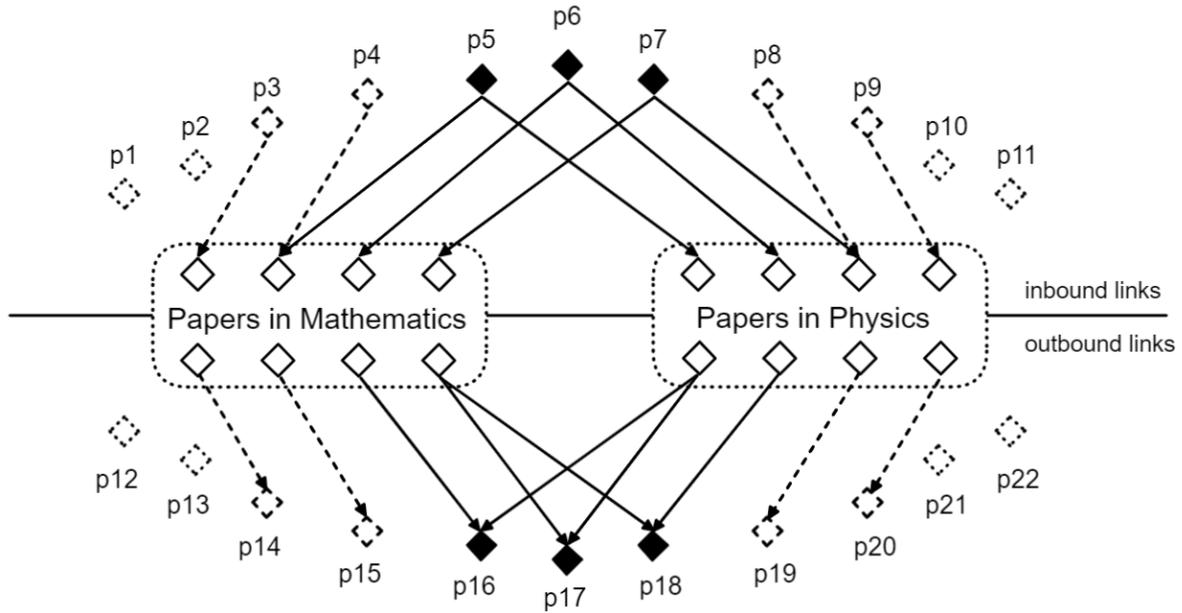

Fig. 1 Collections of papers in 'Mathematics' and 'Physics', solid diamonds represent intersections of papers with inbound links or outbound links.

We extend Google distance[8,9] to measure distance between different collections rather than single nodes. We call this collection-based distance, *knowledge distance*. It can be calculated as following:

$$kd(x,y) = \frac{log(max(|X|,|Y|)) - log(|X \cap Y|)}{log(|G|) - log(min(|X|,|Y|))}$$

Here, $x$ and $y$ are two fields whereas $X$ and $Y$ are the collections of all papers linked into $x$ and $y$ via citations, respectively. $G$ is the entire MAG and it can be sliced according to publication years, for example, $G_y$ represents the set of all papers up to the year $y$, and $P_y$ represents all papers published in the year $y$. $G_{y-1} = G_y - P_y$. Generally, $0 <= kd(x,y) <= 1$, in case $kd(x,y) > 1$, it is set to 1. The knowledge relevance or closeness between $x$ and $y$ can be defined as $kr(x,y) = 1 - kd(x,y)$. Inbound citations of all papers can be calculated based on outbound citations as following:

$$C_j = \bigcup_{i=1}^{n} c_{ij}$$

where $C_j$ denotes the total inbound citations to the paper $j$, and $c_{ij}$ counts citations from the paper $i$ to the paper $j$.

We have calculated knowledge distances between all top and secondary fields of MAG. The result is plotted over years from 1955 to 2020 (Fig. 2). Fig. 2a shows the knowledge distance between the top field 'Mathematics' and others. Kant declared "a doctrine of nature will contain only as much proper science as there is mathematics capable of application there." [11] D'Arcy Thompson believed that the criterion of true science lay in its relation with mathematics[15]. Cohen has studied the interaction between the natural science and the social science[3], and held that the social science is "softer" than the natural science. If we take closeness to 'Mathematics' as a measure of hardness of a field, then 'Computer Science' and 'Physics' are the hardest as shown in Fig. 2a. 'Biology' has always been the softest in STEM. The distance between 'Mathematics' and 'Biology' is even farther than that between 'Mathematics' and 'Political Science'. Other than 'Computer Science', 'Engineering' and 'Physics', almost all fields are moving away from 'Mathematics' over time. Eugene Wigner's famous saying "unreasonable effectiveness of mathematics in the natural sciences" [12] is wisely revised by Hamming as "unreasonable effectiveness of mathematics in computer science, engineering and physics" [13].

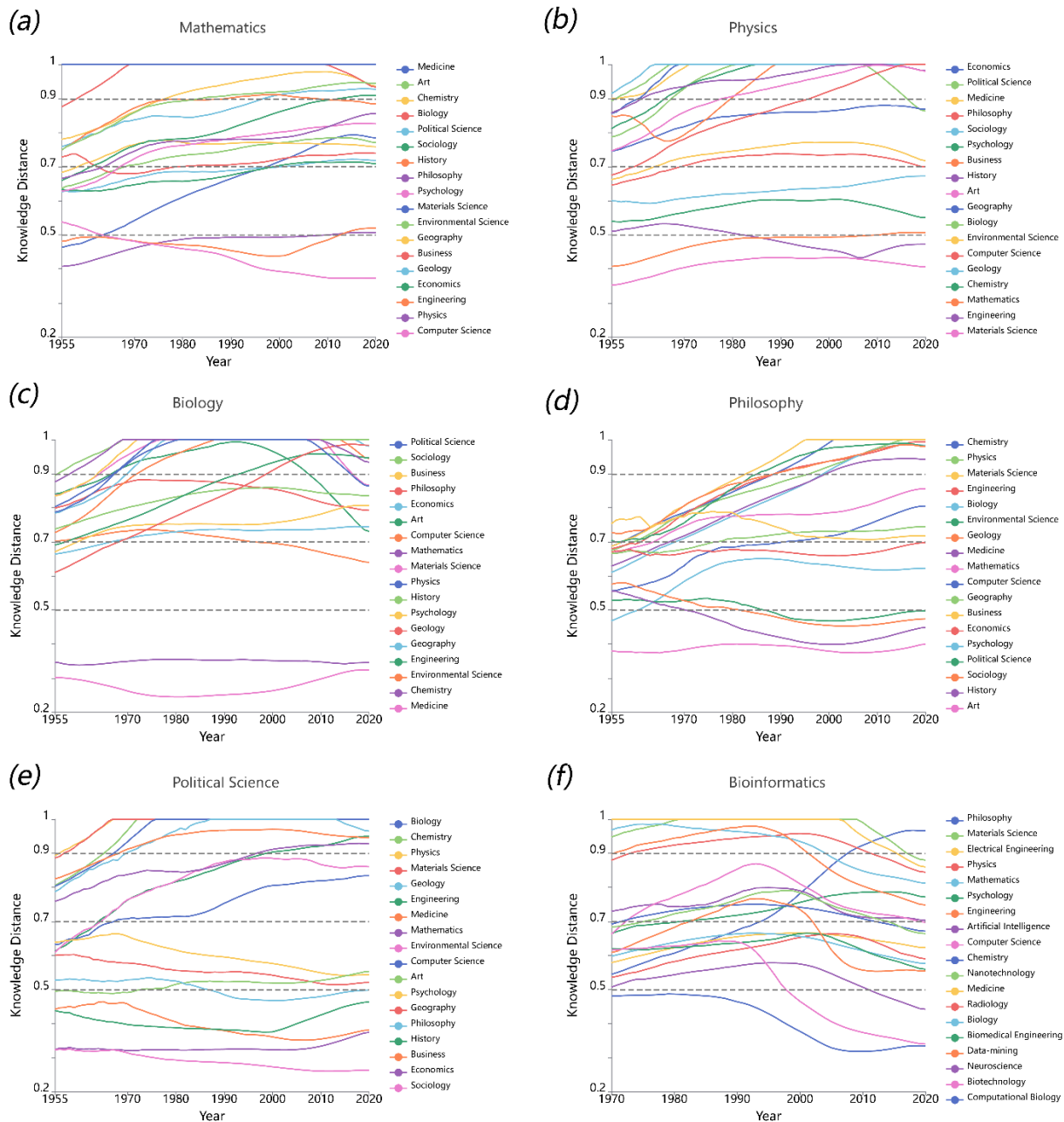

Fig. 2 Knowledge distances between academic disciplines. In order to have all papers being treated uniformly, books and patents are deleted as well as papers that have no inbound references, and more than 70 million papers are processed.

Fig. 2b illustrates the distance between 'Physics' and other disciplines. Before the year 2000, 'Physics' was mostly the farthest from 'Biology' and 'Medicine', which corroborates respective statements by Mayr and Arbesman[14,16] that physics and biology are two distinct kinds of thinking. 'Mathematics' and 'Engineering' are always closest to 'Physics'. Other STEM fields are generally close to 'Physics' but literary fields are moving away from it.

'Biology', 'Chemistry' and 'Medicine' always cluster together (Fig. 3c) but they tend to stay away from all other fields. The fact that 'Biology' is far from 'Mathematics' can also be circumstantially evidenced by the argument whether mathematics is essential to science between biologist Wilson and mathematician Frankel[18,19]. 'Engineering' began first in the year 2000, then, both 'Mathematics' and 'Computer Science' join in 2010 to approach 'Biology'. If we expand the 'Biology' tree to look at its secondary subfields, e.g., 'Bioinformatics', it is clear why 'Biology' is becoming harder (Fig. 2f). Note that 'Data-mining' and 'Biomedical Engineering' are subfields of 'Computer Science' and 'Engineering', respectively. These interdisciplinary links hold 'Biology' closer to the harder parts of STEM.

It is obvious that all fields in humanities and social sciences stay closely (Fig. 3d and Fig. 3e), and physical scientists and engineers are least concerned about them.

Each field has two or three tightly coupled fields. For examples, 'Mathematics' is closest to 'Computer Science' and 'Physics'; and 'Biology' is closest to 'Chemistry' and 'Medicine'. Apart from tightly coupled fields, fields generally tend to be farther and farther over time. The relationship between fields can be divided into four types by the knowledge distance: Family ($kd<=0.5$), Neighbor (0.5-0.7), Stranger (0.7-0.9), and Alien (>0.9). For example, 'Mathematics', 'Computer Science' and 'Physics' are mostly located in one family. If A and B are in one Family, B and C are in another Family, then A and C are at least in the same Neighbor with few exceptions. Fields within same Family keep close and stable relationship while fields in a Neighbor are getting farther. A STEM field has fewer family members than any literary field. The trend of separation is stronger in Stranger and Alien.

**Knowledge space, knowledge clusters and knowledge consilience threshold.**

The degree of consilience can be defined by knowledge distance. The shorter the distance, the closer the consilience. 19 top fields are expanded to get 293 secondary fields. A knowledge space graph can be constructed in which a node is secondary field and the weight of an edge is the knowledge distance between two secondary fields. A minimum spanning tree (MST)[20] is a subset of connected edges with minimal total weight. The MST of the knowledge space is essentially its backbone. The edges with the minimum distance can be added to MST gradually until the average degree of the graph reaches 10 for better visualization as shown in Fig. 3a. Each top field is given a distinct color. The size of a secondary field is the number of papers it contains. The subfields of 'Chemistry' and 'Engineering' are not clustered as tightly as others but rather scattered around the entire knowledge space.

When the knowledge distance between two top fields is less than a given value, *kct* (knowledge consilience threshold), we consider them unified under the *kct*. We use *cluster* to refer to a collection of unified fields with a certain *kct* value. When *kct* is gradually increased, the number of clusters will be reduced, and eventually reach a single unified cluster, the ideal state of Wilson's consilience.

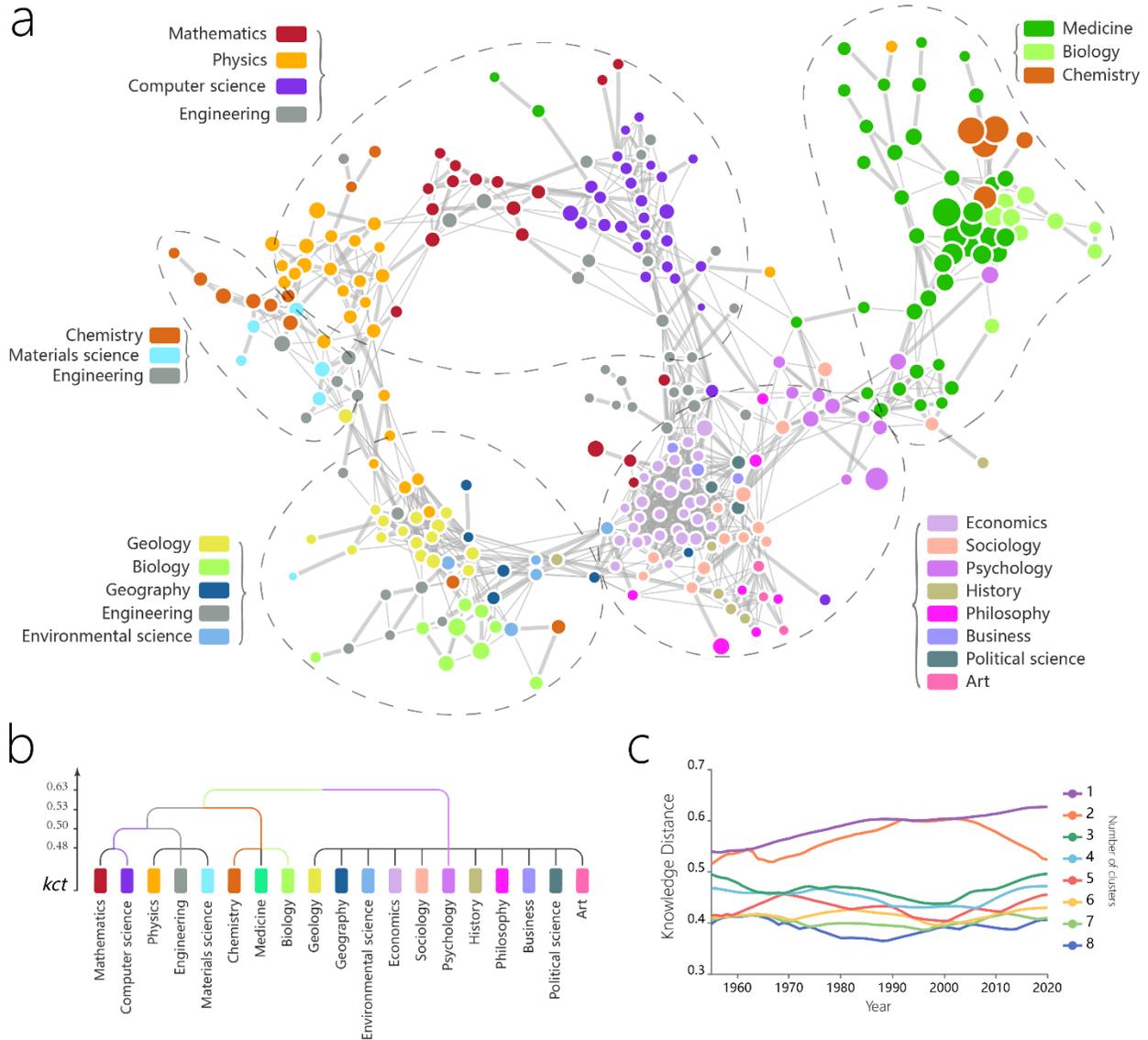

Fig. 3 Knowledge space, knowledge clusters and the knowledge consilience threshold. Knowledge distance is calculated based on data from MAG of 2020. (a) knowledge space graph. (b) knowledge cluster formation. (c) knowledge consilience threshold changes over time.

Fig. 3b shows how clusters are formed as *kct* changes. When *kct* is 0.4, the knowledge space is divided to 9 clusters. When *kct* is further increased to 0.48, the number of clusters is reduced to

4, and STEM is divided into three clusters, namely, (Biology, Chemistry, Medicine), (Mathematics, Computer Science) and (Physics, Material Science, Engineering) while the entire literary side is already unified. The composition of these four clusters remains stable, except that before the year 1978, when 'Computer Science' was located within 'Physics' cluster rather than 'Mathematics'. This shows that 'Computer Science' may have become more mathematical as it matures.

When *kct* is further increased to 0.5, the number of clusters is reduced to 3, the cluster (Biology, Chemistry, Medicine) is still independent, while the rest of STEM is merged together. This is different from Kagan's three culture division: natural sciences, social sciences and humanities[4]. *Kct* for (Biology, Chemistry, Medicine) to hold together is very low, ranging from 0.34 to 0.35, therefore, the cluster is stable.

When *kct* reaches 0.53, STEM is unified, and two cultures emerge. Perhaps geologists may feel somewhat disappointed that they are put on the softer bank. The grand unification threshold is 0.63, where all disciplines are consolidated via a fragile link between 'Medicine' and 'Psychology'. The literary culture has a generally very low threshold and therefore easy to be unified while STEM undertakes much higher unification threshold.

The consilience of entire knowledge is becoming hopeless as *kct* of almost every cluster increases with time (Fig. 3c). A notable exception is when the number of clusters is 2, and *kct* declines after 2003 as 'Chemistry' and 'Engineering' become closer due to interdisciplinary activities which make STEM slightly tighter, and easier for (Biology, Chemistry, Medicine) cluster to merge with the rest of STEM.

**Conclusions.** The clusters formed by MST in knowledge space is consistent with the cluster composition determined by *kct*. Knowledge distance and *kct* are effective metrics for knowledge evolution. Branches of human learning tend to couple tightly with other branches that are near while remote branches are farther and farther. The dream of grand unification might just be a nostalgia.

**References**

1. Snow, C.P. *The Two Cultures*, The REDE Lecture 1959, reissued edition (Cambridge University Press, 2011).
2. Labinger, J.A. & Collins, H. *The One Culture? A Conversation about Science* (University of Chicago, 2001).
3. Cohen, I.B. *Interactions: Some Contacts between the Natural Sciences and the Social Sciences* (MIT Press, 1994).
4. Kagan, J. *The Three Cultures: Natural Sciences, Social Sciences and Humanities in the 21st Centuries*, (Cambridge University Press, 2009).

5. Wilson, Edward O. *Consilience: The Unity of Knowledge*, (Pantheon Books, 1999).
6. Sinha, A., Shen, Z., Song, Y., Ma, H., Eide, D., Hsu, B., & Wang, K. An Overview of Microsoft Academic Service (MAS) and Applications. *Proceedings of the 24th International Conference on World Wide Web* (WWW '15 Companion). ACM, New York, NY, USA, 243-246. DOI=http://dx.doi.org/10.1145/2740908.2742839 (2015).
7. Wang, K., et al, A Review of Microsoft Academic Services for Science of Science Studies, *Frontiers in Big Data*, vol.2 December (2019).
8. Cilibrasi, R.L. & Vitanyi, P.M.B. The Google Similarity Distance. *IEEE Transactions on Knowledge and Data Engineering* **19**(3), 370-383 (2007).
9. Milne, D., & Witten, I.H. An effective, low-cost measure of semantic relatedness obtained from Wikipedia links. *Proceedings of the AAAI 2008 Workshop on Wikipedia and Artificial Intelligence* (2008).
10. Petersen, M., et al. Methods to account for citation inflation in research evaluation. *Research Policy* **48**, 1855–1865 (2019)
11. Kant, I. *Metaphysical Foundation of Natural Science* (Cambridge University Press, 2011).
12. Wigner, E. P. The unreasonable effectiveness of mathematics in the natural sciences, *Comm. Pure Appl. Math.*, **13**, Feb. (1960).
13. Hamming, R. W. The Unreasonable Effectiveness of Mathematics. *The American Mathematical Monthly*, **87**(2) 81-90 (1980).
14. Mayr, E. *The Growth of Biological Thought* (Belknap Press, 1982).
15. Thompson, D. *On Growth and Form* (Dover Publications, 1992).
16. Arbesman, S. *Over Complicated: Technology at the Limits of Comprehension* (2016).
17. Jones, B. The Burden of Knowledge and the 'Death of the Renaissance Man': Is Innovation Getting Harder? *The Review of Economic Studies*, Vol. 76, No. 1 (2009).
18. Wilson, Edward O. Great Scientist ≠ Good at Math, *Wall Street Journal*, April 5, (2013)
19. Frankel, Edward Don't Listen to E. O. Wilson, *Slate*, April 9, (2013)
20. Kruskal, J. B. On the shortest spanning subtree of a graph and the traveling salesman problem. *Proceedings of the American Mathematical Society*. **7** (1): 48–50, 1956.